\documentclass[a4paper,11pt]{article}
\pdfoutput=1 

\usepackage{jinstpub} 
\let\ifpdf\relax
\usepackage{color}
\usepackage[normalem]{ulem}
\usepackage{lineno}
\usepackage{graphicx}
\usepackage{subcaption}
\usepackage{mwe}
\usepackage{amsmath}
\usepackage{gensymb}
\usepackage[symbol]{footmisc}
\usepackage{siunitx}
\usepackage{newtxmath,newtxtext}

\newcommand{\wirecell}{\textit{Wire-Cell}\xspace}
\newcommand{\jinstref}{ref.\xspace}

\def\Put(#1,#2)#3{\leavevmode\makebox(0,0){\put(#1,#2){#3}}}

\newlength{\figwidth}
\newlength{\fighalfwidth}
\title{\boldmath \center \LARGE Augmented Signal Processing in Liquid Argon Time Projection Chambers with a Deep Neural Network
}

\author[a,1]{H. W. Yu, \note[1]{Corresponding author}}
\author[a]{M. Bishai, }
\author[a]{W. Q. Gu, }
\author[b]{M. F. Lin,}
\author[a]{X. Qian, }
\author[b]{Y. H. Ren,}
\author[a]{A. Scarpelli,} 
\author[a]{B. Viren, }
\author[a]{H. Y. Wei,}
\author[c]{H. Z. Yu, }
\author[b]{K. Yu,}
\author[a]{C. Zhang}

\affiliation[a]{Physics Department, Brookhaven National Laboratory, Upton, NY, USA}
\affiliation[b]{Computational Science Initiative, Brookhaven National Laboratory, Upton, NY, USA}
\affiliation[c]{Sun Yat-Sen (Zhongshan) University, Guangzhou, China}

\emailAdd{hyu@bnl.gov}

\abstract{
The Liquid Argon Time Projection Chamber (LArTPC) is an advanced neutrino detector technology widely used in recent 
and upcoming accelerator neutrino experiments. It features a low energy threshold and high spatial resolution 
that allow for comprehensive reconstruction of event topologies. In current-generation LArTPCs, the recorded data consist of 
digitized waveforms on wires produced by induced signal on wires of drifting ionization electrons, which can 
also be viewed as two-dimensional (2D) (time versus wire) projection images of charged-particle trajectories. For such 
an imaging detector, one critical step is the signal processing that reconstructs the original charge projections from the 
recorded 2D images. For the first time, we introduce a deep neural network in LArTPC signal processing to 
improve the signal region of interest detection. By combining domain knowledge (e.g., matching information 
from multiple wire planes) and deep learning, this method shows significant improvements over traditional methods. 
This work details the method, software tools, and performance evaluated with realistic detector simulations.
}


\begin{document}
\maketitle
\flushbottom

\section{Introduction}\label{sec:intro}

The Liquid Argon Time Projection Chamber (LArTPC) is a key detector technology for many current and anticipated future accelerator neutrino 
experiments~\cite{Acciarri:2016smi, Abi:2017aow, Antonello:2015lea, Abi:2020evt}. When charged particles traverse
through the LAr medium, both scintillation light and ionization electrons are created. While the detection of scintillation 
light provides the event time, the detection of ionization electrons affords high-resolution position and energy information 
about particle trajectories. The rich event topology information provided by the LArTPC offers unique advantages in performing 
electron-photon separation. Together with superior calorimetry capability, the LArTPC is an excellent detector to 
study $\nu_\mu$ to $\nu_e$ oscillations, which may hold the key to answering some remaining questions in the 
neutrino sector~\cite{Diwan:2016gmz}.

\begin{figure}[thb]
  \centering
  \includegraphics[width=1.0\figwidth]{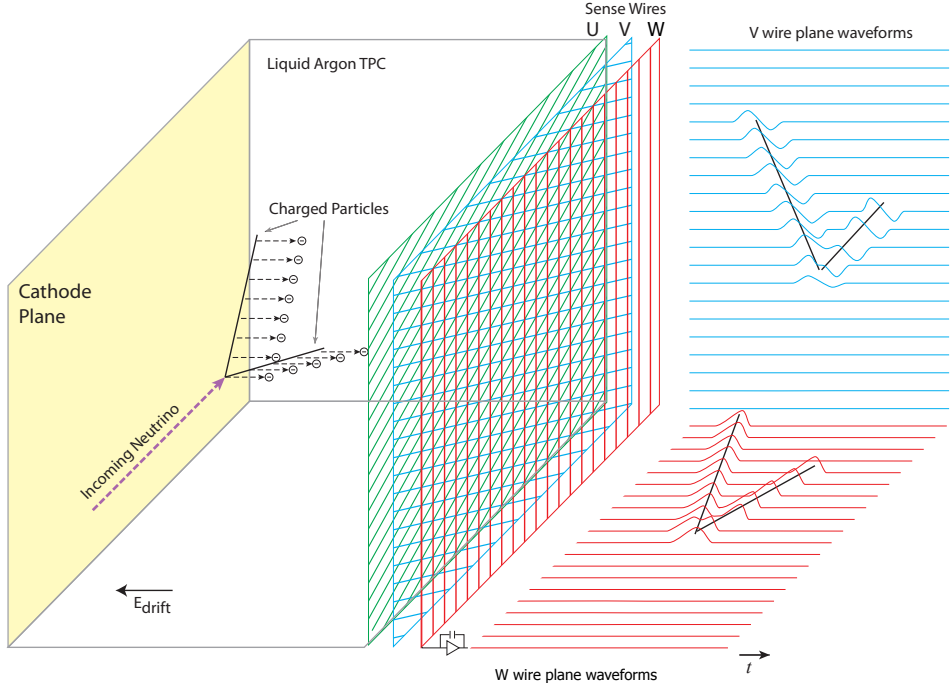}
  \caption{Diagram from ref.~\cite{Acciarri:2016smi} showing conceptual configuration of a typical three wire plane LArTPC and illustrating the signal formation of it. The signal in the U induction plane is omitted from the diagram for simplicity.}
  \label{fig:lartpc-concept}
\end{figure}

The current-generation LArTPCs are typically equipped with wire readouts, instead of pixelated readouts that is being developed rapidly~\cite{Dwyer:2018phu}, under the 
consideration of cost and heat production of electronics inside LAr~\cite{Radeka:2011zz}. In this configuration, the anode 
typically consists of multiple (commonly three) wire planes with different wire orientations. Under a uniform external electric field, the 
produced ionization electrons drift toward the anode planes at a (known) constant speed. The arrival time of ionization 
electrons, combined with the event time provided by the light detection system, allows for a position reconstruction of 
activities along the electric field direction.
Figure~\ref{fig:lartpc-concept} ~\cite{Acciarri:2016smi} illustrates the signal formation in a typical LArTPC configuration. 
The combination of wire signals from different planes provides the position information 
perpendicular to the electric field. Together, three-dimensional (3D) reconstruction of activities inside LArTPC can be made. To realize 
this concept, wire planes with different orientations must record signals from the same ionization electron cloud multiple times. 
This is achieved using both induction and collection wire planes. By properly configuring the electric fields between 
wire planes, the ionization electrons can fully pass through the first few (induction) wire planes then get collected by 
the last (collection) wire plane. When ionization electrons move close to the wires, induced currents can be detected.
As governed by Ramo's theorem~\cite{ramo}, the induced current has bipolar and 
unipolar shapes on the induction-plane and collection-plane wires, respectively. While signal reconstruction
for the collection-plane wires is generally simpler, it is more complex for the induction-plane wires 
given the potentially large cancellation effect of the bipolar signals. In this work, we introduce a novel signal 
processing procedure utilizing modern deep learning techniques and assisted by correlating information from multiple 
planes. This new procedure leads to an improved reconstruction of ionization electron distributions, which further 
enhances the quality of overall event reconstruction.  

The recorded digitized TPC signal is a convolution of the distribution of the arriving ionization electrons and the 
impulse detector response: 
\begin{equation}\label{eq:conv}
  M(t,x) = \int_{-\infty}^{\infty}\int_{-\infty}^{\infty} R(t-t',x-x') \cdot S(t',x')dt' dx' + N(t,x),
\end{equation}
where $M(t,x)$ is a measurement, such as an Analog to Digital Converter (ADC) value at a given sampling time and wire position; $R(t-t',x-x')$ is the impulse 
detector response, including both the field response that describes the induced current by a moving ionization electron and the 
electronics response from the shaping circuit; $S(t',x')$ is the charge distribution in time and space of the arriving 
ionization electrons; and $N(t,x)$ is the electronics noise. The goal of TPC signal processing is to reconstruct the original 
charge distribution $S$ from the measurement $M$ given the known detector response $R$ in the presence of the electronics 
noise $N$. 

\begin{figure}[thb]
  \centering
  \includegraphics[width=1.0\figwidth]{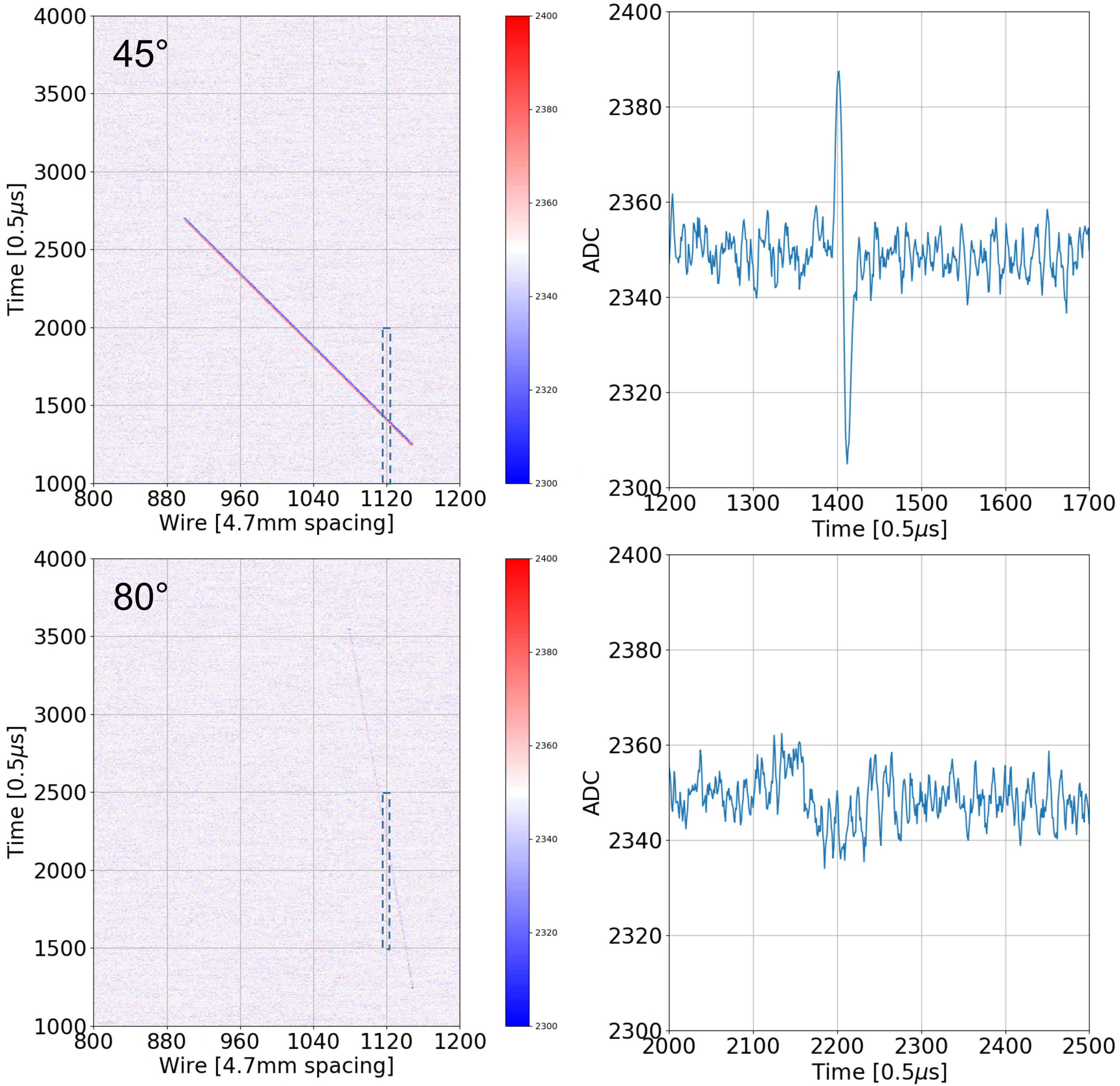}
  \caption{Left: projection of a minimum ionizing particle (MIP) track on one induction plane; 
    right: a section of waveform of wire 1120 from the left plot 
   indicated by the dashed box. Track angle projected on this plane is 45\degree~(Top) and 80\degree~(Bottom) with 
respect to the wire orientation. The recorded signal for the 80\degree~ track is weak as the result of the cancellation 
effect of the bipolar field response and the extended (in time) distribution of ionization electrons.}
  \label{fig:angle-wave}
\end{figure}

Developed from earlier work in \jinstref~\cite{Baller:2017ugz}, the current state-of-the-art algorithm to reconstruct the original 
ionization electron distribution is the so-called ``2D deconvolution'' technique~\cite{Adams:2018dra,Adams:2018gbi}
using Discrete Fourier Transformation. This TPC signal processing algorithm is implemented in the \wirecell
Toolkit~\cite{wirecell_toolkit} and can be run as a plug-in of the LArSoft software suite~\cite{larsoft}. 
Although this algorithm has shown good performance and has been adopted in many experiments, there is still room to improve, 
especially for large angle tracks with respect to the wire orientation. These tracks are known as the 
``prolonged tracks'' because they induce current on a given wire over an extended period of time~\cite{Adams:2018dra}. 
Figure~\ref{fig:angle-wave} illustrates the challenge in signal processing for prolonged tracks (with an 80\degree track 
angle). In this work, the track angle is defined by the track projection on a 2D plane, where the x-axis is the 
wire pitch direction and y-axis is the electron drifting direction. The track angle is measured by the angle between 
track projection and the x-axis in this plane, defined as $\theta_{xz}$. This angle definition is the same as that used in figure~8 of 
\jinstref~\cite{Adams:2018dra}. 

While the TPC signals for a track with a $\theta_{xz} = 45\degree$ are clear (see. figure~\ref{fig:angle-wave}), the ones for a track 
with $\theta_{xz} = 80\degree$ are rather weak. Such a big difference with respect to the track angle is the result of the bipolar
and long-range nature of the induced current for a point source on the induction-plane wires. Thus a prolonged track leads 
to large cancellation of the induced current, resulting in a low signal-to-noise ratio. The same conclusion can be 
drawn when viewing from the frequency domain. First, the average response function of the induction-plane wires must approach to 
zero at low frequency (e.g., figure~12 of~\cite{Adams:2018dra}) as the net charge collected on an induction-plane wire is 
zero~\cite{ramo}. Second, the signal of a prolonged track mostly resides at low frequency. Therefore, the measured signal 
strength, which is the product of signal strength from a prolonged track and the average response function, can become small 
compared with the low-frequency electronics noise, leading to a poor signal-to-noise ratio. To achieve good 
performance of the signal processing, \jinstref~\cite{Adams:2018dra} introduced a region of interest (ROI) detection technique, 
whose goal is to define an ROI in the time domain to contain the distribution of ionization electrons. 
The ROI significantly reduces low-frequency 
noise for the induction planes, which, in turn, increases the signal-to-noise ratio. Obviously, to maximize the 
signal-to-noise ratio, the ROI fully containing the distribution of ionization electrons in the time domain should be as small as possible. For this reason, the ROI finding algorithm in 
\jinstref~\cite{Adams:2018dra} worked on the deconvolved waveform instead 
of the raw waveform because the ROI found in the latter case is typically bigger as a consequence of the extended detector 
response. With a set of heuristic logic, the ROI searching algorithm in \jinstref~\cite{Adams:2018dra} further utilizes the 
connectivity information of ROIs on adjacent wires in a wire plane to enhance ROI detection efficiency as the TPC signals tend to be 
continuous for any charged particle track. Of note, while the ROI detection is crucial for induction-plane signal 
processing, it is less important for collection-plane signals, where the field response is generally unipolar. 

However, even with the usage of ``2D deconvolution'' technique and ``connectivity'' information, some TPC signals, e.g. 
prolonged tracks, are still hard to identify. Adding more information progressively into the heuristic logic of
the ROI detection algorithm~\cite{Adams:2018dra} is not a scalable approach.  Hence, we propose a novel LArTPC signal 
processing procedure based on the Deep Neural Network (DNN). DNNs are known to manage complicated 
correlations and cover solution phase space more efficiently. In the field of high energy physics, there are 
many successful applications of DNNs (see refs.~\cite{Aurisano:2016jvx,Sirunyan:2017ezt,Shimmin:2017mfk,Sadowski:2017ilo,uboone_dnn_paper,Baldi:2018qhe,Radovic:2018dip}, among others). Furthermore, inspired by the \wirecell imaging 
concept in \jinstref~\cite{Qian:2018qbv}, the TPC signals from other wire planes can enhance ROI detection
efficiency by using the geometry information from multiple wire planes. Combining machine learning and domain knowledge, the new DNN 
LArTPC signal processing shows significantly improved performance, especially for prolonged tracks.

This paper is organized as follows: section~\ref{sec:sigproc} briefly reviews the current state-of-the-art signal processing, and section~\ref{sec:method} introduces the new DNN signal processing. The DNN architecture, \wirecell imaging concept, simulation and data pre-processing, and network training are described in sections~\ref{subsec:dnn}, 
~\ref{subsec:geom}, ~\ref{subsec:data-pre-process}, and~\ref{subsec:training}, respectively. The DNN signal processing performance evaluation is presented in section~\ref{sec:results} followed by the section~\ref{sec:summary} summary.

\section{LArTPC Signal Processing}\label{sec:sigproc}

\begin{figure}[thb]
  \centering
  \includegraphics[width=0.9\figwidth]{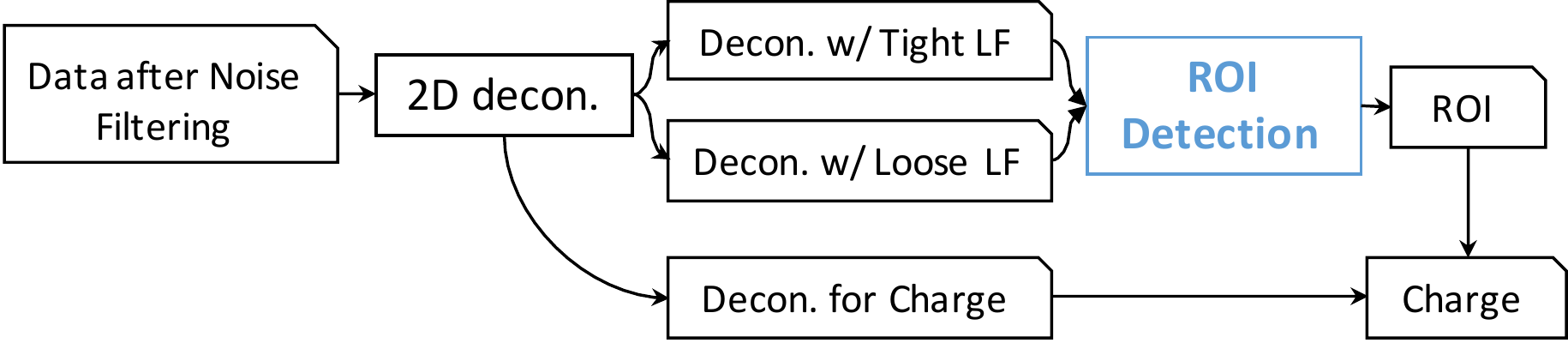}
  \caption{Flowchart of the LArTPC signal processing algorithm in \jinstref~\cite{Adams:2018dra}. 
  ``LF'' stands for low-frequency software filter, while ``decon.'' denotes deconvolution.}
  \label{fig:current-sp-flow}
\end{figure}

As introduced in section~\ref{sec:intro}, LArTPC signal processing extracts ionization charge distribution as a function
of drift time and wire number from the induced current on the readout wires. Figure~\ref{fig:current-sp-flow} shows a simplified 
flowchart of the state-of-the-art signal processing algorithm in Refs.~\cite{Adams:2018dra,Adams:2018gbi}.

\noindent Starting from the noise-filtered waveform~\cite{noise_filter_paper}, several 2D deconvolutions are performed 
with different software filters in the frequency domain. Each deconvolution follows Eq.~\ref{eq:decon}, resolving the 
signal $S$ using measurement $M$ and detector response $R$ in the frequency domain. A follow-up Inverse Fourier Transform 
($IFT$) converts the signal $S$ back to the time and space domain. Software filters ($F$) in the frequency domain are used 
to suppress noise at high and/or low frequencies. Here, the low-frequency filter is necessary (not used) to process 
signals in the induction-plane (collection-plane) wires as the bipolar nature of induction-plane 
wires' field response suppresses signals at low frequency. The different low-frequency filters have distinct features in the signal
extraction. For example, a tight low-frequency filter limiting the bandwidth at a higher frequency value leads to a 
high-purity but low-efficiency signal extraction. On the other hand, a loose low-frequency filter expanding the bandwidth
to lower frequency results in a low-purity but high-efficiency signal extraction.

\begin{equation}\label{eq:decon}
  S(\omega_{t},\omega_{x}) \sim \frac{F(\omega_{t},\omega_{x})\cdot M(\omega_{t},\omega_{x})}{R(\omega_{t},\omega_{x})} \quad \underrightarrow{\quad IFT \quad} \quad S(t,x).
\end{equation}

\noindent With the deconvolved waveform, ROI detection is performed to identify signal regions that are more likely 
caused by ionization signals (instead of noise). The ROI detection step intends to open windows just large enough to 
contain signals and maximize the signal-to-noise ratio. A set of heuristic logic is implemented to detect ROIs 
from the deconvolved signals with loose and tight low-frequency filters based on the connectivity information. With the 
ROI windows defined, the waveforms outside are suppressed to zero, which reduces data size significantly and makes the reconstruction steps that follow computationally more efficient. The detected ROIs are then applied on the deconvolved waveform without 
applying low-frequency filters ("Decon. for Charge" in figure~\ref{fig:current-sp-flow}). While removing the 
low-frequency filters minimizes the distortion on the ionization electron distribution, applying ROIs in the 
time domain keeps the signal-to-noise ratio high, which is necessary for induction-plane wires. 

The goal of ROI detection is to achieve high efficiency and high purity. This can be challenging when the signal-to-noise ratio 
is low, such as with the prolonged tracks described in section~\ref{sec:intro}. In addition, the heuristic logic in ROI 
detection sometimes may fail in a busy situation where numerous activities are present (e.g., near neutrino interaction vertex 
or inside an electromagnetic shower). Section~\ref{sec:method} introduces a novel ROI detection algorithm using a convolutional neural 
network based on information from multiple wire planes, which notably improves signal processing performance.

\section{Deep Neural Network  ROI Detection}~\label{sec:method}

In this section, we describe the details the Deep Neural Network (DNN) architecture, use of geometric 
information from multiple wire planes, the simulation and data pre-processing, and the network training.

\subsection{Neural Network Architecture}~\label{subsec:dnn}

\begin{figure}[thb]
  \centering
  \includegraphics[width=1.0\figwidth]{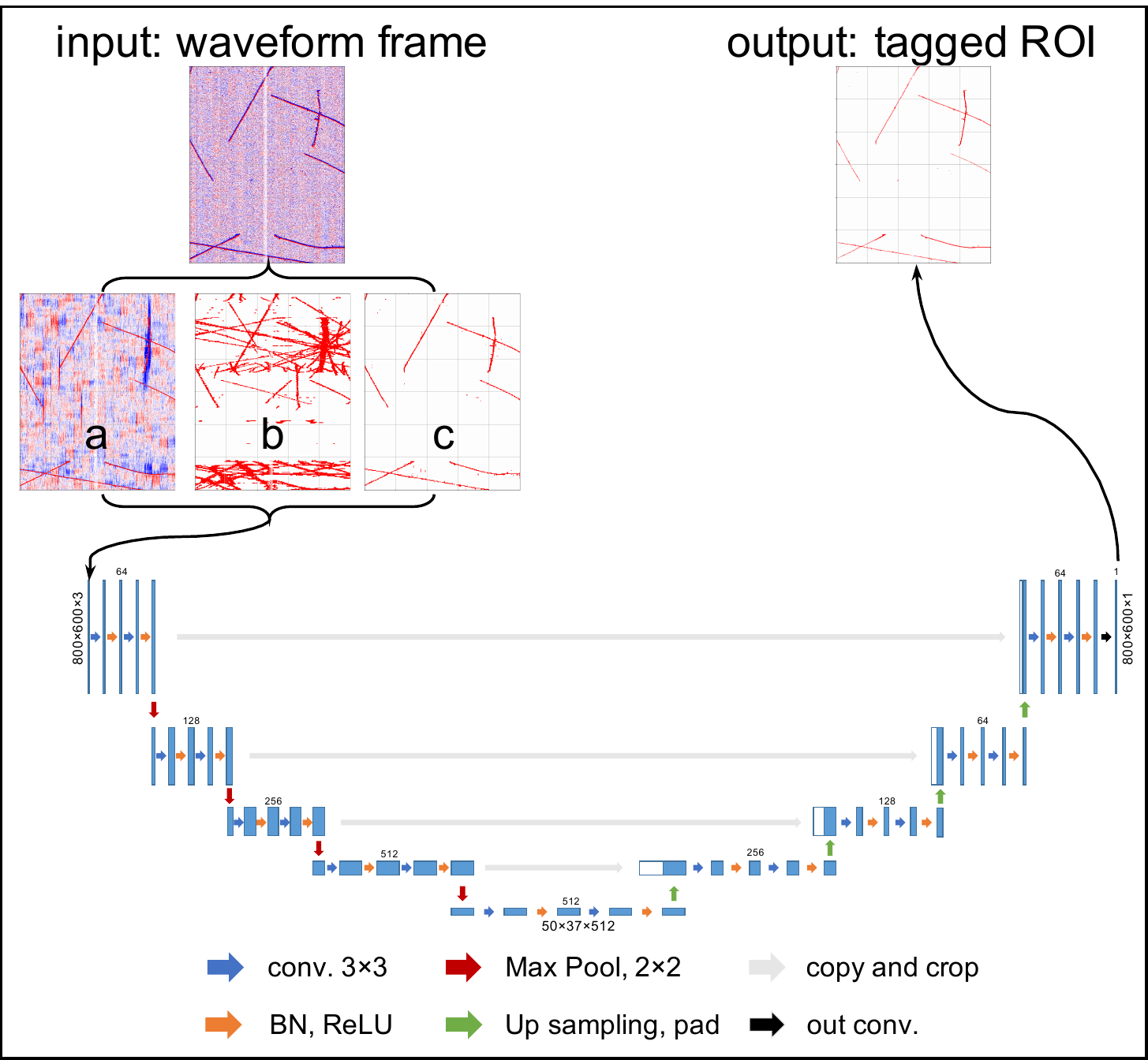}
  \caption{U-Net architecture used in the ROI detection for the DNN LArTPC signal processing. Intermediate 2D 
images are generated from original images containing the raw waveform. Several of these intermediate images are stacked to a 
multi-channel 2D image serving as the U-Net input. Numbers of intermediate images can vary. In this example, 
three are used: a) deconvolved signals from a loose low-frequency filter, b) MP2, and c) MP3. Output of the U-Net 
is a single channel 2D image labeling each pixel as signal 
(i.e., inside ROI) or not.}
  \label{fig:net}
\end{figure}

To apply the DNN, the ROI detection problem is essentially labeling each pixel in a 2D image 
(with one dimension spanning LArTPC readout channels and the other drift time) 
as ``signal'' or ``not-signal'' (noise). Such a problem of dividing pixels from a picture into 
separated groups belongs to the class of machine learning procedures called \textit{semantic segmentation}.
We adopted the U-Net network architecture, introduced by O. Ronneberger et al.~\cite{unet_paper} in 2015, which is a type of convolutional network~\cite{10.1162/neco.1989.1.4.541}. The U-Net has been developed into a family of network architectures (see Refs~\cite{nestedunet_paper, uboone_dnn_paper}, 
among others). Figure~\ref{fig:net} shows the network configuration, which consists of an initial 
encoding path, a final decoding path, and skip connections which bridge these first two paths at various levels.
Each level contains two 3$\times$3 convolution layers and two batch normalization layers. In the encoding path, 
output tensors from each level are downsampled into the next level through a 2$\times$2 max pooling operation.
In the decoding path, output tensors from each level are upsampled by a bilinear upsample operation.
The upsampled tensors are concatenated with tensors copied from the same level of the encoding path to form 
input for each level of the decoding path. The encoding path finds patterns at different levels based on the 
original information from input channels. In the decoding path, output channels are reconstructed from the 
patterns found from deeper levels or the skip connections. In the application of ROI detection, we use
a five-level network. After a factor of 2 condensing in the height and width dimensions of each downsampling operation, in the deepest level 
of the network, pattern recognition is effectively performed in regions of $48 ( = 3 \times 2^4) \times 48$ 
pixels.

\subsection{Using Geometry Information}\label{subsec:geom}

\begin{figure}[thb]
  \centering
  \includegraphics[width=1.0\figwidth]{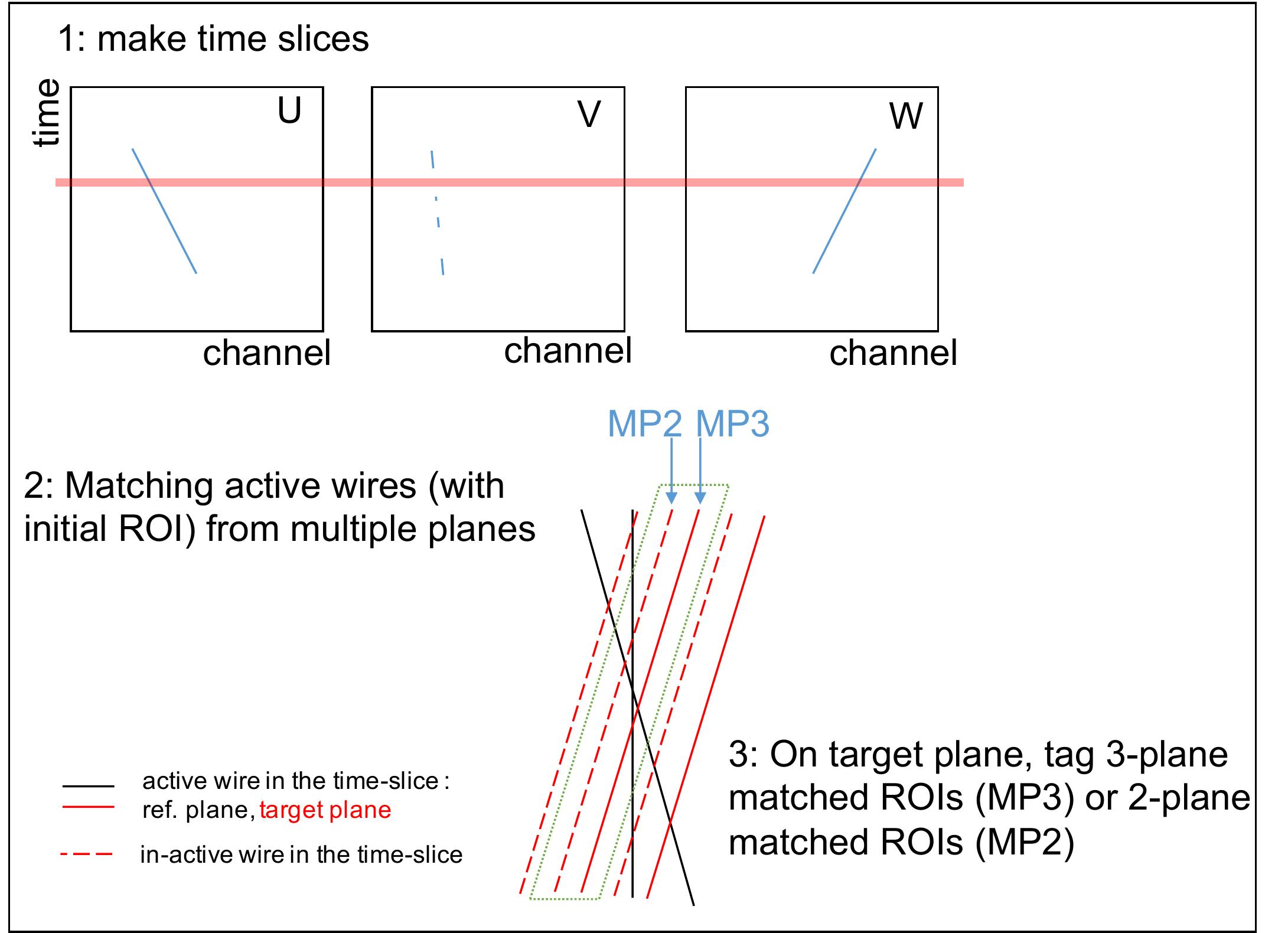}
  \caption{Illustration of creating MP2 and MP3 images. The application of geometric constraints is performed 
at each time slice (four time ticks).}
  \label{fig:mp-projection}
\end{figure}

In a LArTPC with multiple wire planes, each ionization electron is independently sensed by each wire plane. 
Therefore, information from the other wire planes can be used to assist ROI detection in the targeted 
wire plane by correlating the geometric relation among wires with different orientations. Such geometric
constraints are expected to be more efficient for LArTPCs with three or more wire planes.
For example, we use a three-plane configuration with the first two induction planes, labeled as ``U'' and ``V'', and the final collection plane, denoted as ``W''.
This idea of \textit{multi-plane constraint} in signal processing is inspired by the \wirecell tomographic reconstruction technique from \jinstref~\citep{Qian:2018qbv}.
The procedure for implementing geometric 
constraints on the induction-plane wires is described below and illustrated in figure~\ref{fig:mp-projection}:
\begin{enumerate}
\item For each channel, initial signal ROIs are formed by combining the deconvolved signals with tight 
and loose low-frequency software filters (see figure~\ref{fig:current-sp-flow}). The combination uses
the connectivity information but skips the majority of heuristic logic in refining ROIs. 
\item Across the channels, these initial signal ROIs are sliced into contemporaneous time slices of 
fixed duration (e.g., four time ticks or two $\mu$s). This choice ensures negligible loss of information
following the Nyquist sampling theorem~\cite{Nyquist:1928zz}.
\item In a given time slice, we determine a subset of channels from each plane consisting of those that 
are inside the initial signal ROIs.
\item For each induction-plane channel inside the subset, we determine if any of its wires overlap with two wires 
from the other two subsets (one from each).\footnote{In the case of wrapped wires, one channel may correspond to multiple 
wires.} The successfully matched channels are referred to as 
the \textit{three-plane coincidence} (MP3).
\item For each induction-plane channel outside of the subset, we determine if any of its wires overlap with two wires 
from the other two subsets (one from each). The successfully matched channels are referred to as 
the \textit{two-plane coincidence} (MP2).
\item Steps 3 to 5 are repeated for every time slice. 
\end{enumerate}

\noindent Identifying the coincidence (MP2 or MP3) is a combinatorial problem with the potential to require 
prohibitive computational expense. To combat this, we developed an optimization technique for the primitive operations 
used to determine wire overlap by exploiting the symmetries of uniform wire direction and pitch. More details can be 
found in appendix~\ref{sec:raygrid}.

As shown in figure~\ref{fig:net}, three 2D images are used as input to the DNN to detect the ROI for each induction wire plane. 
The first image is the deconvolved result after applying the loose low-frequency software filter. The content in each 
2D pixel (one channel and one time tick) is a float number representing the reconstructed ionization charge. The second
image is the result of MP2. The content in each pixel is a number with the Boolean type and unity (zero) labeling the 
pixel to be inside (outside) the MP2. The last image is the result of MP3 also with Boolean type. Although MP2 and MP3
are obtained after rebinning four time ticks into one time slice, it is straightforward to determine if one pixel with 
one time tick is inside the MP2 or MP3 region. Figure~\ref{fig:example-input-label} depicts an example of these arrays.

\subsection{Simulation, Data Pre-processing, and Truth Label Preparation}\label{subsec:data-pre-process}

The data samples used to train the network are generated from a comprehensive and detailed simulation over 
multiple stages of physics. These begin with an initial sampling of kinematics from cosmic ray muon 
models~\cite{Engel:2018akg}; particle passage through detector material and the resulting energy 
depositions~\cite{Geant4}; the production of electrons, including recombination effects~\cite{larsoft};
and detector signal and noise simulation~\cite{wirecell_toolkit}. The detector response in the 
last step is provided by the \wirecell Toolkit that includes ionization electron drift, diffusion, 
stochastic fluctuations, field response, and electronics response.
The data pre-processing steps include excess noise filtering, initial deconvolution, and initial ROI finding.
Excess noise filtering is particularly crucial in dealing with real detector data, but it is less important for simulation. 
The initial deconvolution and ROI finding are discussed in section~\ref{subsec:geom} and more details can be 
found in \jinstref~\cite{SP1_paper}. The simulation and data pre-processing chain is validated with the
ProtoDUNE Single Phase detector (described in \jinstref~\cite{pdune_sp_performance}).

\begin{figure}[thb]
  \centering
  \includegraphics[width=1.0\figwidth]{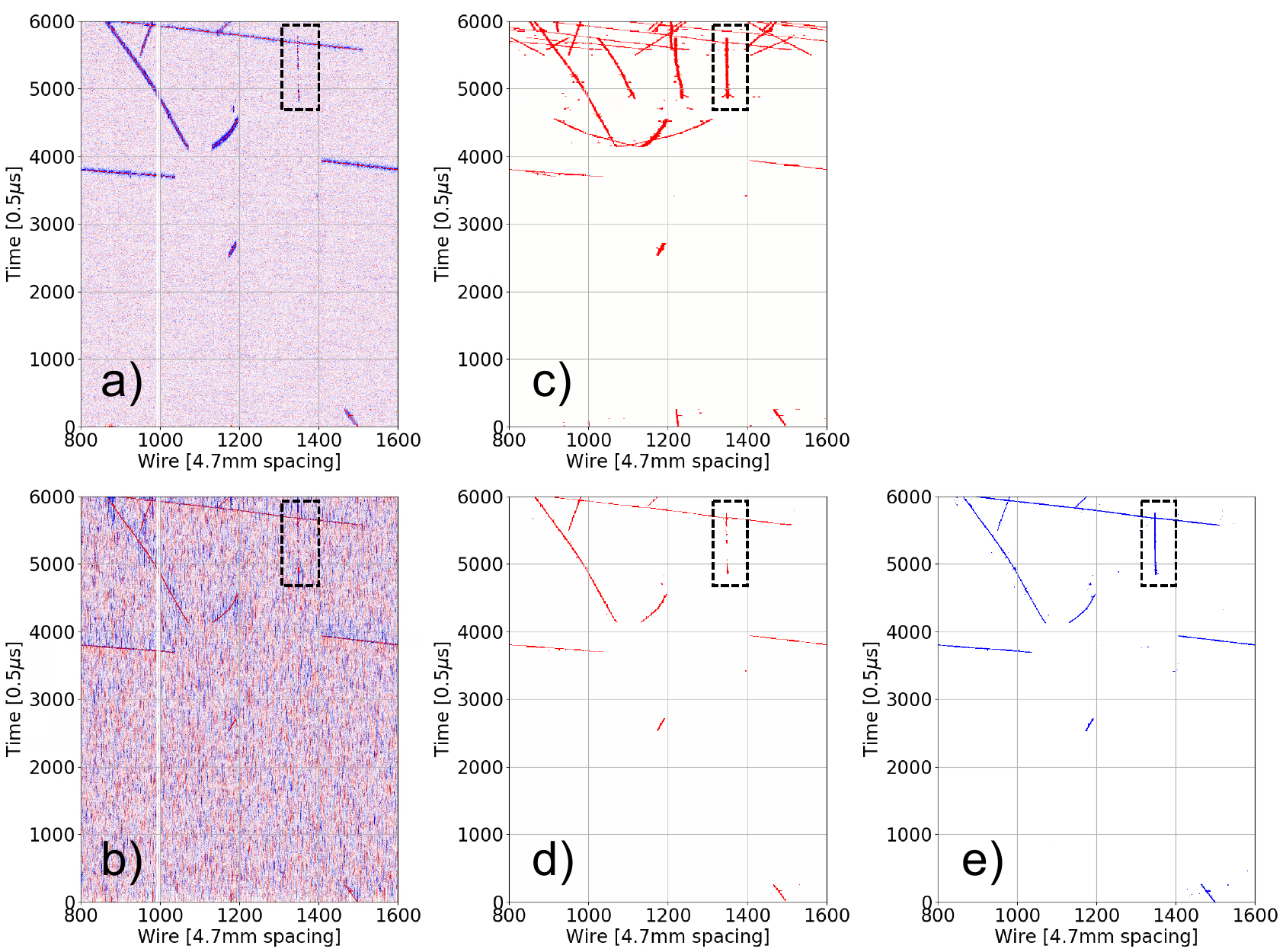}
  \caption{Example neural network input channels obtained from cosmic ray simulation for an induction wire plane:
  a) deconvolved image with a tight low-frequency filter; 
  b) deconvolved image with a loose low frequency filter; 
  c) multi-plane 2-plane coincidence (MP2); 
  d) multi-plane 3-plane coincidence (MP3);
  and e) truth information. Here, the deconvolved images with tight and loose low-frequency filters from different 
wire planes are used to produce the MP2 and MP3 images. The displayed images feature original binning
of 6000 time ticks and 800 channels. A prolonged track is indicated by the black rectangle dashed line. The input to DNN consists of b), c), and d). }
  \label{fig:example-input-label}
\end{figure}

The input to DNN consists of MP2 (figure~\ref{fig:example-input-label}c), MP3 (figure~\ref{fig:example-input-label}d), 
and deconvolved signals from a loose low-frequency filter (figure~\ref{fig:example-input-label}b).
Notably, MP2 and MP3 are generated from the deconvolved signals with tight (figure~\ref{fig:example-input-label}a) and loose low-frequency filters.
For each event sample, there are a total of 6000 time ticks covering 3~ms drift time. 
The number of channels are 800, 800, and 960 for induction U, induction V, and collection W planes, respectively. 
The average number of cosmic tracks is about 11 per event.
To limit memory use, 10 time ticks are rebinned into one time bin, which limits the typical data size of the induction plane to 600 time bins and 
800 channels.
For any 2D image, the content value in each rebinned pixel is taken as the average of the original value in 10 pixels (e.g., number of electrons for the 
deconvolved signal and 0 or 1 for MP2 and MP3).
While the number of ticks used in this rebinning process could be further optimized, such a rebinning choice has minimal impact on the ROI detection because a majority of the ROIs ($>$95\%) have a length longer than 10 time ticks.
For network training purposes, the \textit{truth} label is created by applying a simple ideal-detector model to the initial simulated 
ionization electron distributions (a smearing with Gaussian distribution). A threshold is applied on the charge 
of a rebinned pixel (one channel and 
one time bin covering 10 time ticks) to determined if a rebinned pixel contributes to an ROI. In particular, the 
threshold is chosen to be 100 electrons on the average charge per time tick in a rebinned pixel.
This value is much lower than the equivalent noise charge (ENC) from the entire electronics readout chain, which typically exceeds 400 electrons per 
time tick. The average number of (true) ROIs in an induction plane is about 3500 per event. 
Figure~\ref{fig:example-input-label}(e) shows an example of the truth label. 

\subsection{Network Training}~\label{subsec:training}

\begin{figure}[thb]
 \centering
 \includegraphics[width=0.9\figwidth]{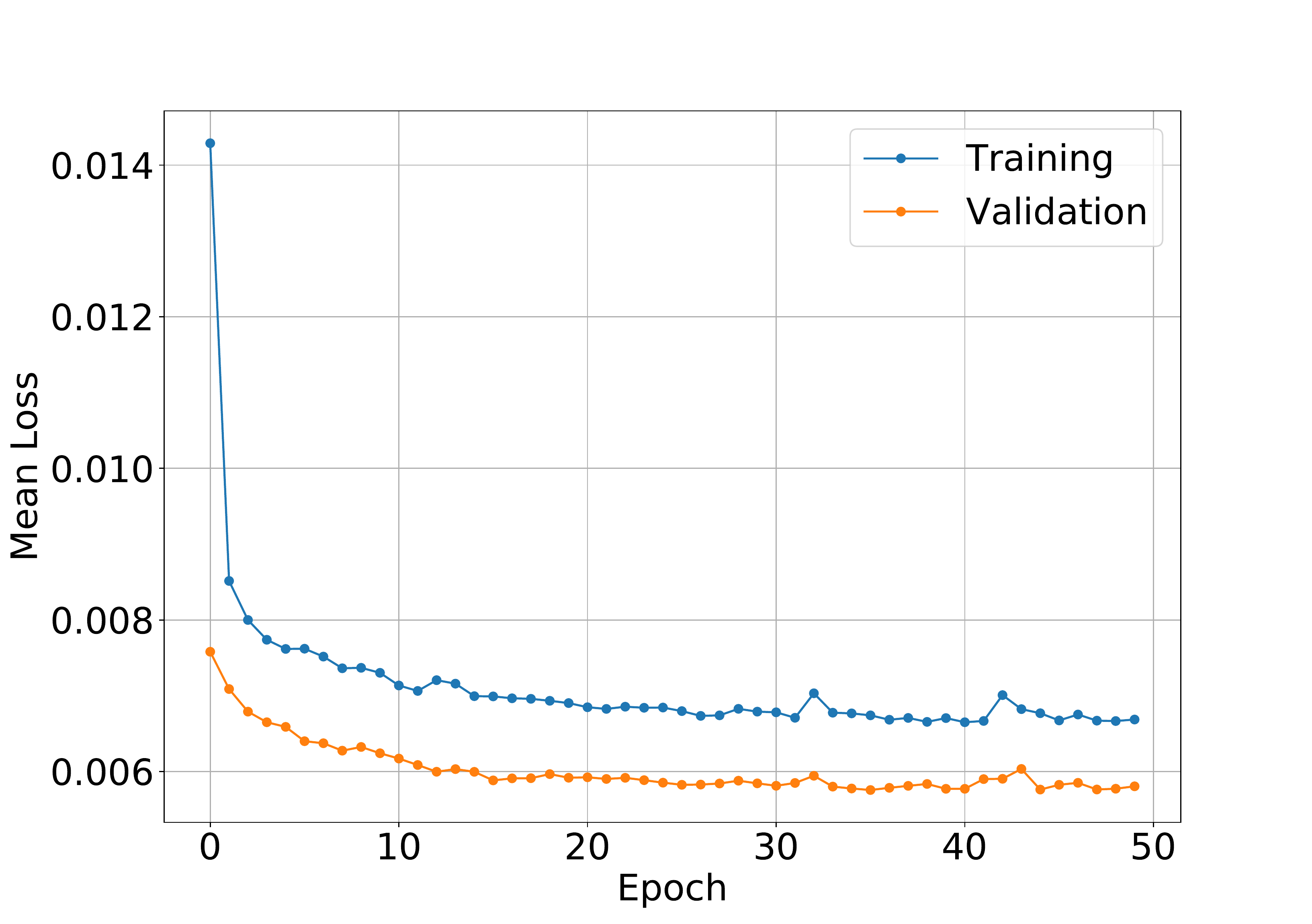}
 \caption{Epoch mean loss as a function of epoch number.}
 \label{fig:loss}
\end{figure}

The network and supporting utilities are implemented using PyTorch~\cite{pytorch} and are publicly accessible in 
\jinstref~\cite{yuhw:unet_pytorch}. We trained the network on a platform with Intel i9-9900K CPU and NVIDIA 2080 Ti GPU 
with 11 GB VRAM~\cite{rtx2080ti}.
The models used in the following paper are trained with 50 epochs.
A training epoch consists of a total of 500 event samples (450 for training and 50 for validation) and requires six minutes on this platform.
Each sample consists of one tensor holding the binary ``signal''/``not-signal'' truth label elements and a second tensor holding three 2D arrays
(deconvolved signal with loose low-frequency filter, MP2, and MP3). With the cosmic ray simulation, each sample contains about 5 to 20 cosmic rays, resulting in an average of 3500 ROIs per sample. We use binary cross-entropy~\cite{pytorch:bce} as the loss 
function and stochastic gradient decent~\cite{pytorch:sgd} with momentum 0.9 and learning rate 0.1 as the optimizer.
Figure~\ref{fig:loss} shows the epoch mean loss as a function of the epoch number for both training and validation samples.
We can see the two loss curves following a consistent trend and flattening out after about epoch 30. 
The absolute difference between the two curves are caused by the data sample difference
between the training and validation datasets.
In the next section, performance of this DNN ROI detection is evaluated.

\section{Performance and Discussions}\label{sec:results}

The performance of the trained network with simulation is initially evaluated by focusing on the more challenging cases 
of the ideal (large-angle) prolonged tracks. For each 3D track, two projection angles, $\theta_{xz}(V)$ and
$\theta_{xz}(U)$ with respect to V and U wire planes, are used to described its 3D direction. 
A larger projection angle with respect to a wire plane means a longer ROI in time, which is more challenging 
to detect. Two metrics are used to evaluate the performance with the rebinned pixels.
Pixel-wise (rebinned) efficiency is defined as:

\begin{equation}\label{eq:pix-eff}
  \textrm{Pixel Efficiency} := \frac{\textrm{\# of correctly label pixels}}{\textrm{\# of ROI pixels from truth label}}.
\end{equation}

\noindent Pixel-wise purity is defined as:

\begin{equation}\label{eq:pix-pur}
  \textrm{Pixel Purity} := \frac{\textrm{\# of correctly label pixels}}{\textrm{\# of ROI pixels from reconstruction}}.
\end{equation}

\begin{figure}[thb]
  \centering
  \includegraphics[width=1.0\figwidth]{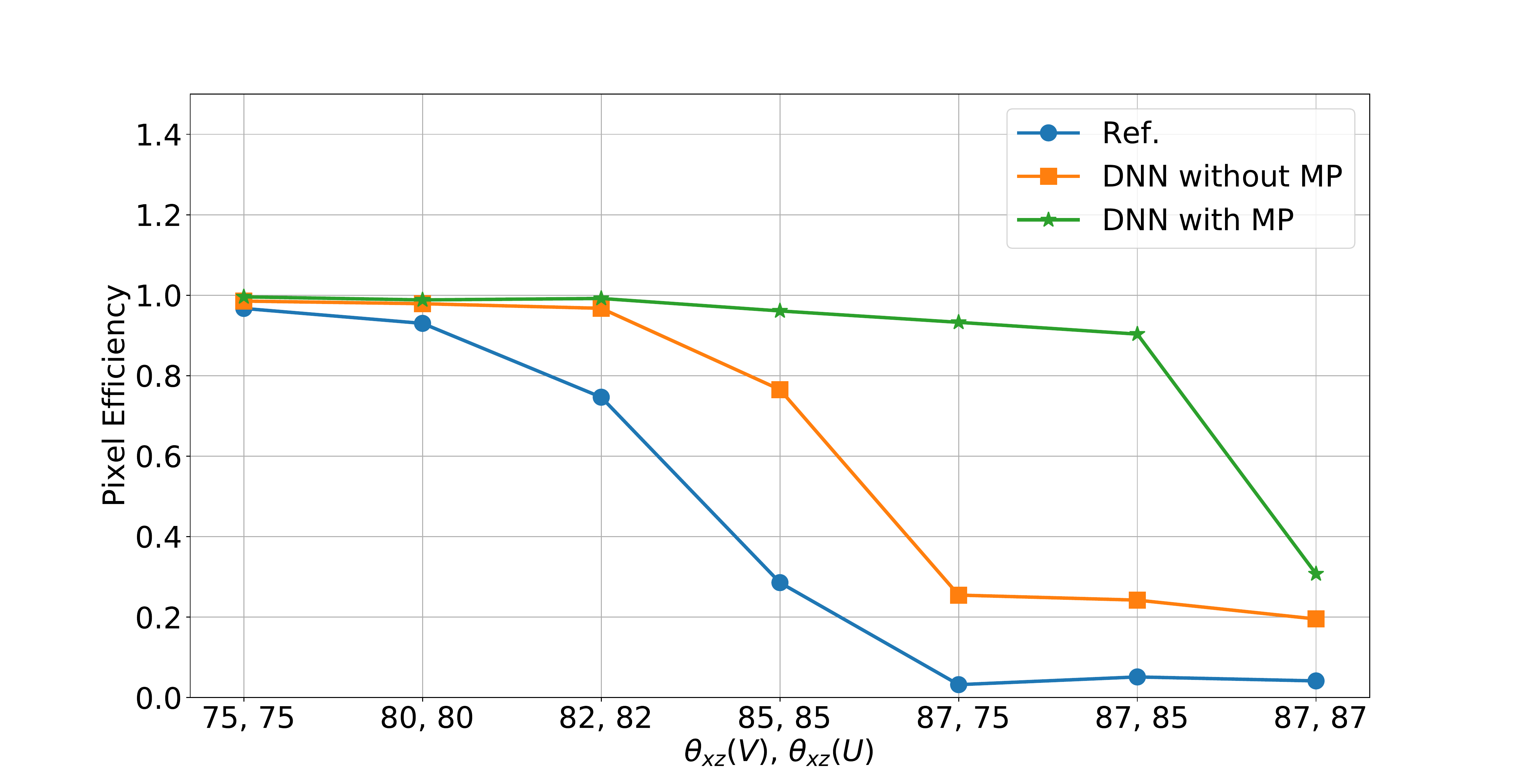}
  \includegraphics[width=1.0\figwidth]{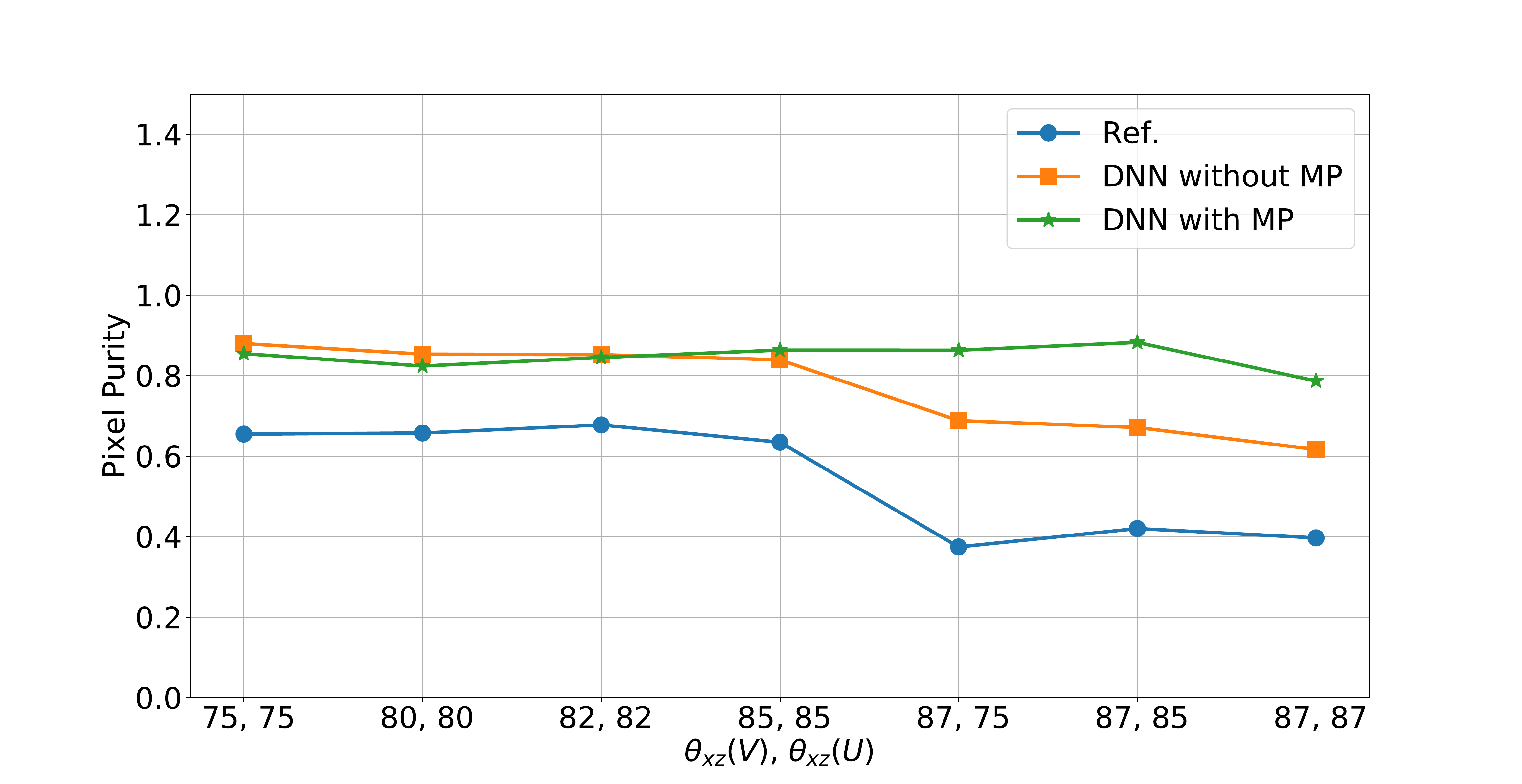}
  \caption{Pixel-wise (rebinned) ROI detection efficiency and purity for three ROI detection algorithms 
evaluated using simulated prolonged tracks. X-axis is track projection angle on the respective induction V and U planes in degrees. ROI detection is performed on the targeted induction V plane.}
  \label{fig:eff-pur}
\end{figure}

\noindent In this evaluation, induction V wire plane is used as an example. Figure~\ref{fig:eff-pur} compares 
pixel-wise ROI detection efficiency and purity for three algorithms: 1) current state-of-the-art ROI detection 
algorithm from \jinstref~\cite{SP1_paper} as the reference, 2) DNN ROI detection without the multi-plane geometry 
information, and 3) DNN ROI detection with the multi-plane geometry information.
Clear improvements can be gleaned from the DNN ROI detection algorithm:
\begin{enumerate}
\item The DNN ROI detection algorithm outperforms the reference algorithm, especially for extreme large-angle 
tracks ($> 85\degree$).
\item DNN ROI without the multi-plane information performs better than the reference but not as good as DNN ROI 
with the multi-plane information.
In particular, by comparing the last three data points with 87\degree ~projection 
angle on the V plane (75\degree, 85\degree, and 87\degree ~projection angles on the U plane), we conclude that
the multi-plane geometry information is more effective when the projection angles on the other wire planes
are small (i.e., less prolonged). In the extreme case when the track travels completely perpendicular to the
anode wire planes, the projection angles become 90\degree ~for both induction planes, and the multi-plane
geometry information is less useful.
\end{enumerate}

\noindent Using various, reconstructed event images, figure~\ref{fig:event-disp} shows additional DNN ROI detection algorithm improvements.
Row (1) of figure~\ref{fig:event-disp} shows the reconstructed images of simulated ideal tracks with 
projection angles 87\degree ~on V plane and 85\degree ~on U plane for the three 
ROI detection algorithms (columns b-d), as well as the truth label (column a).
Improvements in terms of efficiency and purity numbers (shown in figure~\ref{fig:eff-pur}) are evident.
Rows (2) and (3) in figure~\ref{fig:event-disp} 
show reconstructed images from simulations generated from cosmic rays.
While row (2) focuses on the reconstruction of a prolonged cosmic ray track, row 
(3) centers on a busy interaction region of a charged pion. The DNN ROI detection algorithm with 
multi-plane information, column (d), has the best performance among all scenarios 
when compared with truth labels in column (a).
The same conclusion can be drawn when applying the DNN ROI detection algorithm on the ProtoDUNE Single Phase data~\cite{pdune_sp_tdr}. Preliminary results can be found in \jinstref~\cite{yuhw:dnnroi_aps}.

\begin{figure}[thb]
  \centering
  \includegraphics[width=1.0\figwidth]{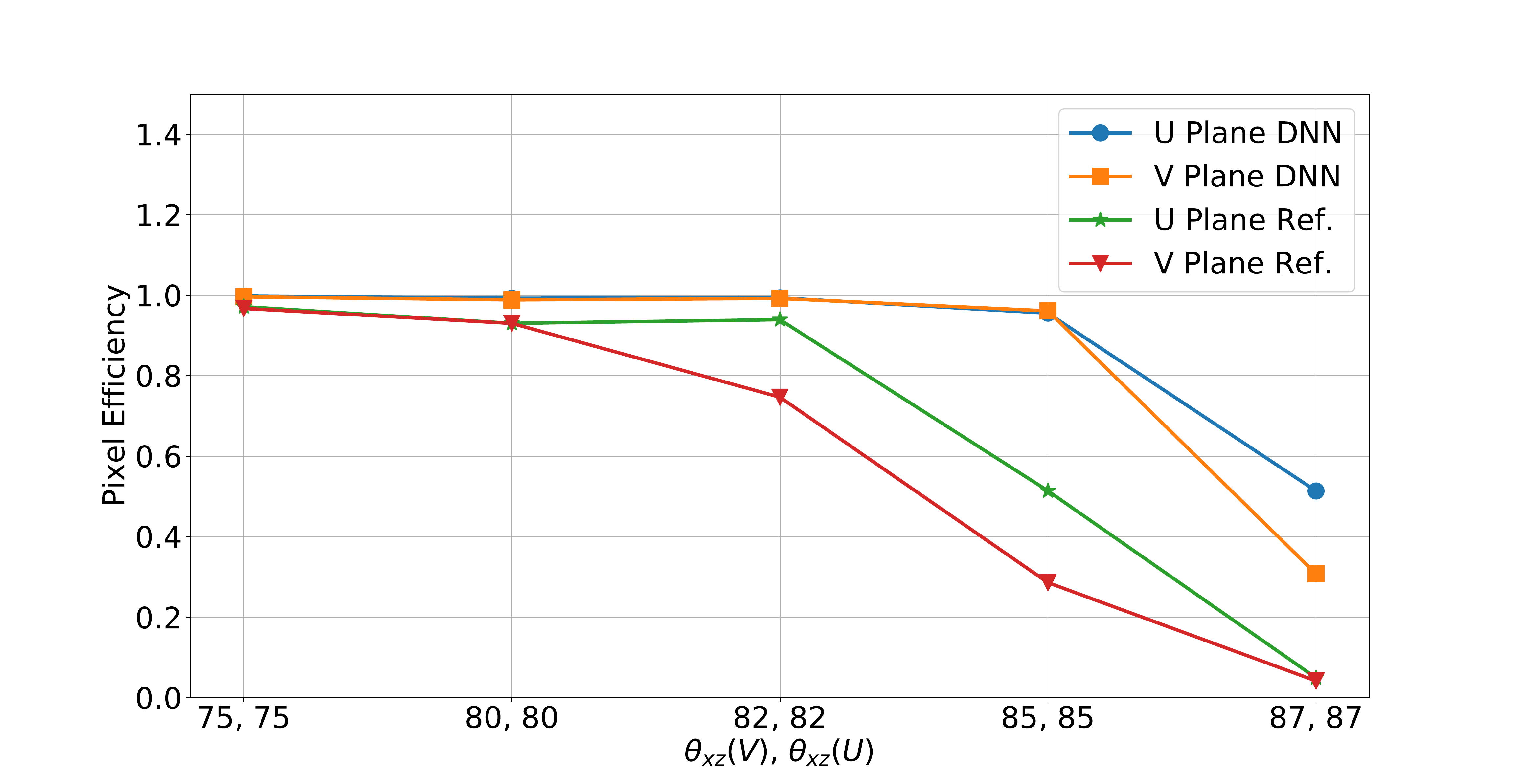}
  \caption{Pixel-wise (rebinned) ROI detection efficiency comparing U and V planes using two algorithms. ``Ref.'': ROI detection algorithm described in \jinstref~\cite{SP1_paper}; ``DNN'': DNN ROI detection with multi-plane information proposed in this paper. X-axis is track projection angle on the respective induction V and U planes in degrees.}
  \label{fig:perf-u}
\end{figure}

\begin{figure}[thb]
  \centering
  \includegraphics[width=1.0\figwidth]{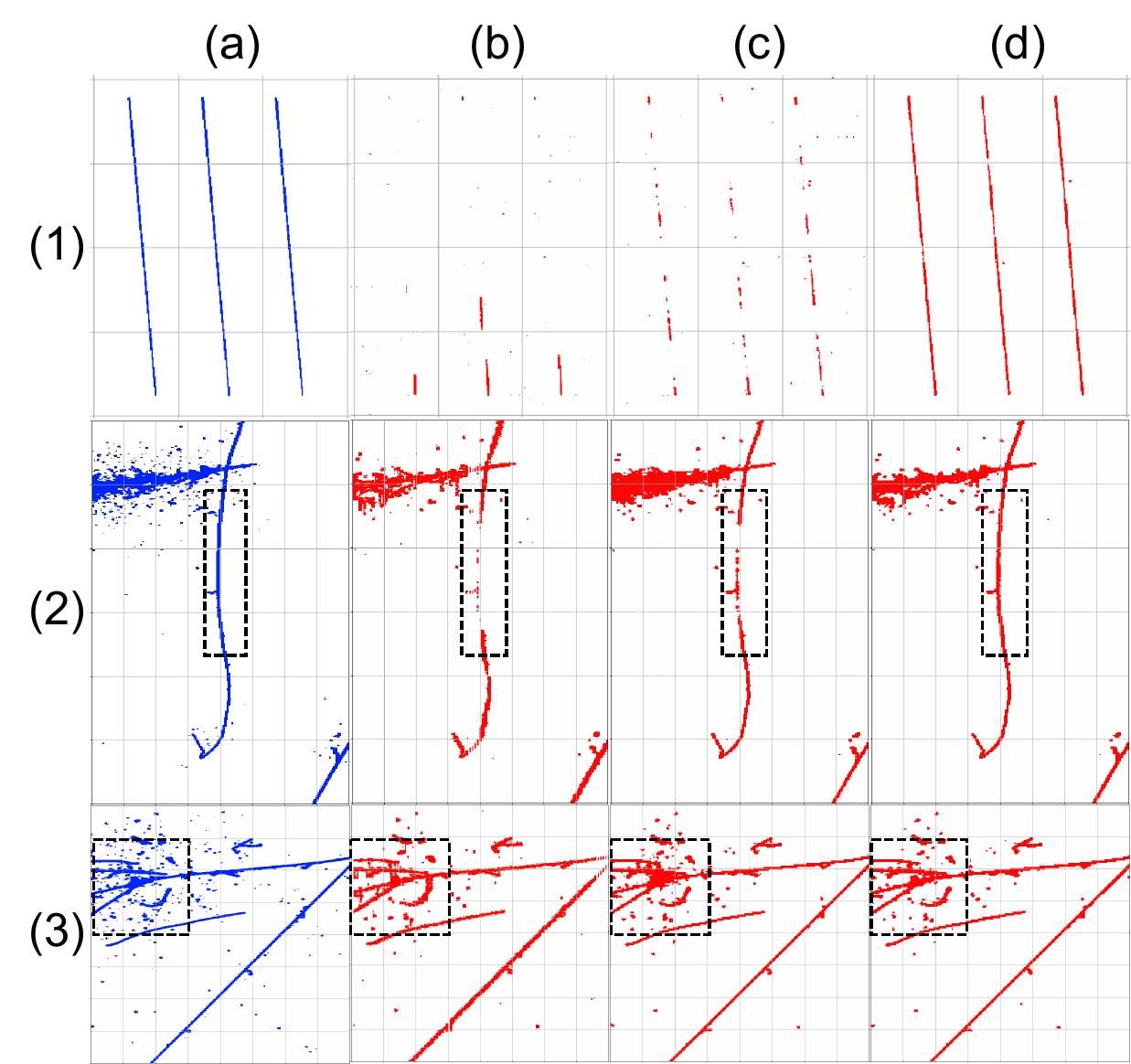}
  \caption{Event display showing the detected ROIs on different types of 
event topologies: (1) Ideal straight-line prolonged tracks with track angle of 87\degree ~projected to V plane and 85\degree ~projected to U plane, (2) cosmic track with a section featuring a large 
projection angle (indicated by the black rectangle dashed line), and (3) an interaction vertex of a charged pion with many activities (indicated by the black rectangle dashed line). 
ROI detection is performed on the V plane (target plane).
The column labels are: (a) truth label, (b) reference ROI detection 
algorithm~\cite{SP1_paper}, (c) DNN ROI detection algorithm without the multi-plane 
geometry information, and (d) DNN ROI detection algorithm with the multi-plane geometry 
information.}
  \label{fig:event-disp}
\end{figure}

Here we show the DNN ROI detection algorithm performance on the most challenging second induction plane (V plane) as an example.
For the collection plane (W plane), given the unipolar signals (therefore negligible cancellation effect on the raw signal), a simple thresholding ROI finding algorithm is found to have a good performance. Therefore, we did not use the DNN ROI finding on the W plane in this work.
The DNN ROI finding has a slightly better performance on the first induction plane (U plane), see figure~\ref{fig:perf-u}. This behavior is because of the slower drift velocity of ionization electrons at the U plane than that at the V plane,  which is the result of the bias voltage setting to ensure the 100\% transparency condition~\cite{doi:10.1139/cjr49a-019}. The slower drift velocity leads to a more extended field response function (in time), which in turn reduces the bipolar cancellation effect on the raw signal and increases the signal-to-noise ratio given a fixed intrinsic electronics noise.

In addition to U-Net, we also implemented and evaluated two other network structures: 
URestNet and Nested-U-Net. URestNet~\cite{uboone_dnn_paper} is U-Net with residual 
connections. Nested-U-Net~\cite{nestedunet_paper} is U-Net with dense skip connections. 
With proper optimizations in the training, all three networks perform similarly. 
U-Net uses slightly less ($\sim$10\%) GPU memory in both training and inferencing.

For a production environment with mainly C++ software stack, we implemented application programming interfaces (APIs) to use 
TorchScript~\cite{torchscript} in the \wirecell Toolkit~\cite{wirecell_toolkit}. The 
trained models are serialized in the TorchScript format. Table~\ref{tab:resource-comp} 
shows a comparison of the inferencing time, memory usage, and VRAM use among the 
three ROI detection algorithms.

\begin{table}[h!]
\caption{Comparison of resource usage for three ROI detection algorithms. Reference is 
the current ROI detection algorithm from \jinstref~\cite{SP1_paper}; 
DNN CPU/GPU: DNN ROI detection algorithm proposed in this paper, 
inferencing with CPU/GPU, respectively.}
\label{tab:resource-comp}
\centering
\begin{tabular}{ | c | c | c | c | }
 \hline
 Method   & Time per plane [sec] & Memory [GB] & VRAM [GB] \\
 \hline
 Reference     & 0.40 & 1.3 & -   \\ 
 DNN CPU  & 16.7 & 4.8 & -   \\  
 DNN GPU  & 0.14 & 3.7 & 3.7 \\
 \hline
\end{tabular}
\end{table}

\section{Summary}\label{sec:summary}

For the first time, we have developed a novel DNN ROI detection algorithm for LArTPC signal 
processing.
As described in this paper, we adopted the ProtoDUNE configuration in the experiment, but the generic idea and procedure should work on all types of projective wire readouts.
The implementation in the \wirecell Toolkit utilizes the TorchScript APIs, which
enables using trained PyTorch models in the C++ production environment. This algorithm takes advantage 
of modern machine learning techniques, as well as the domain knowledge specific to LArTPC signal processing 
(2-D deconvolution, multi-plane geometry correlation, etc.). Quantitative evaluations show this 
DNN ROI detection algorithm outperforms the current state-of-the-art ROI detection algorithm. 
Such improvement is expected to generate a better reconstruction of event topology in a LArTPC, 
particularly for detectors with higher noise and lower signal-to-noise ratio (e.g., with warm electronics~\cite{Antonello:2015lea}). 



\acknowledgments

This work is supported by the U.S. Department of Energy, Office of Science, Office of High Energy Physics and Early Career Research Program under contract number DE-SC0012704. We thank the DUNE collaboration for use of its Monte Carlo simulation software and related tools. We would also like to show our gratitude to Charity Plata (Computational Science Initiative of Brookhaven National Laboratory) for her inputs during the manuscript preparation.

\appendix

\section{Fast Projection using the Ray Grid Technique}\label{sec:raygrid}

Central to the use of geometry information for improving signal processing efficiency, as well as in the \wirecell 3D tomographic reconstruction technique~\cite{Qian:2018qbv}, are a number of primitive operations that must be carried out numerous times by these higher-level algorithms. 
Optimization of these frequently called operations is critical for overall speed. 
Two core and related operations involve locating the intersection point of a wire from each plane and then locating a wire from a third plane relative to this intersection.
The so-called \textit{ray grid} method is used to optimize these operations by exploiting two symmetries of an idealized wire plane: fixed wire pitch and direction. 
Exploiting these symmetries transforms the problem from solving a vector equation to a simpler problem of applying indices.

\begin{figure}[htbp]
\centering
\includegraphics[width=0.9\textwidth,clip,trim=3cm 3cm 3cm 3cm]{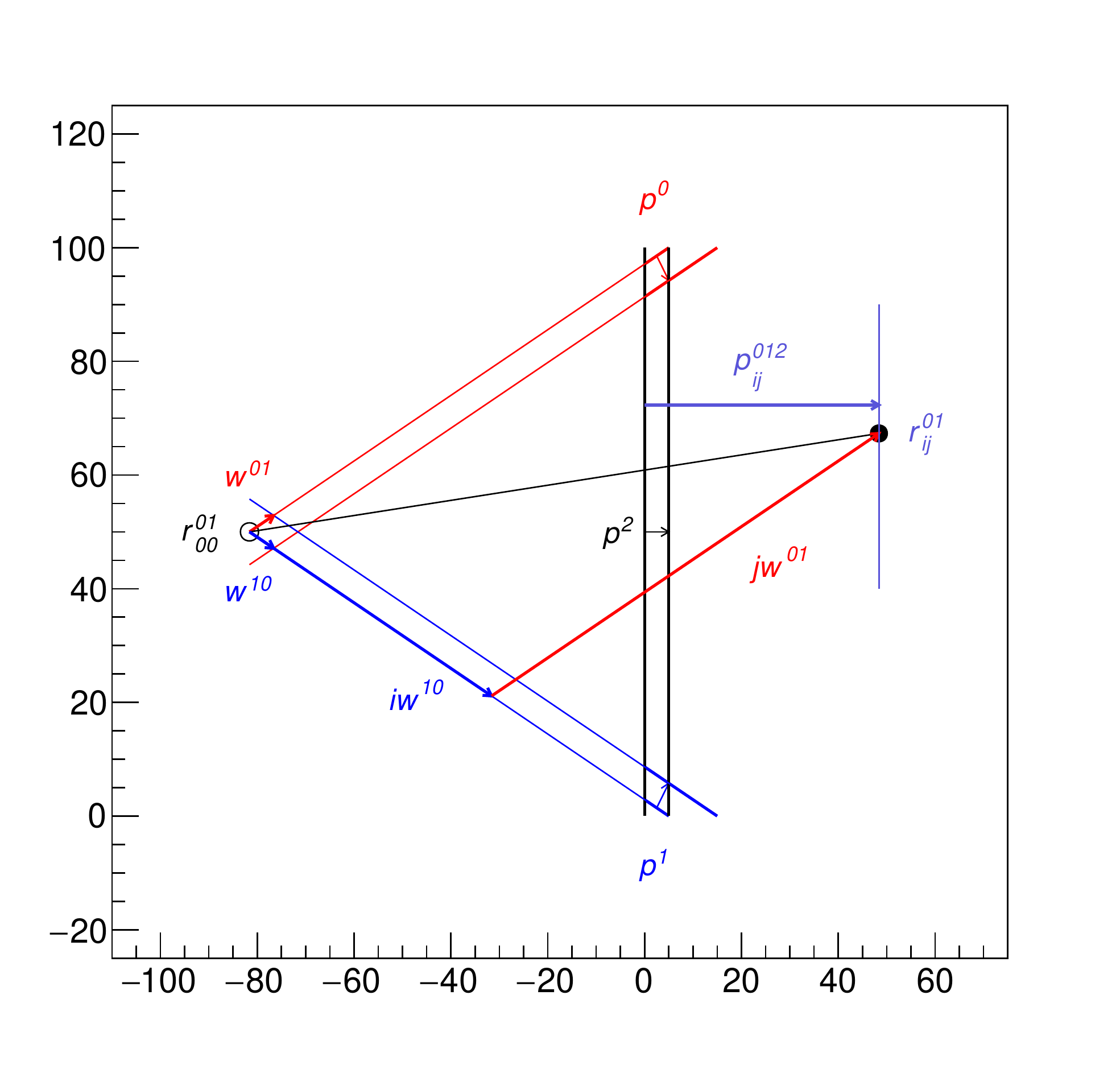}
\caption{\label{fig:raygrid} Example points, vectors and tensors involved in constructing and using ray grids. 
  Layers 0, 1 and 2 are represented in red, blue and black, respectively. 
  See text for further definitions.}
\end{figure}

With these symmetries assumed, we start by generalizing a fixed number of parallel wires in a wire plane to an infinite number of parallel \textit{rays} in a \textit{layer}.
This generalization allows us to define the active area of an anode by introducing two logical layers, one providing horizontal and the other providing vertical boundaries via rays with pitch equal to the active height and width, respectively.
The higher-level algorithms do not care about wires per se but about the logical lines running parallel to and halfway between two neighboring wires and in so in practice we associate the abstract rays to these lines.
Finally, as the higher-level algorithms work in a projected space, we assume all layers are co-planar, which allows the problem to reduce to two dimensions.

A set of rays in a layer and their relationship to rays in a second or third layer may be categorized by a number of tensors listed below and illustrated for a three layer problem in figure~\ref{fig:raygrid}. 
Within one layer we may consecutively number the rays and we mark these tensor indices with subscripts ($\in \{i,j,k\}$). 
We identify a special ray with an index of zero. 
The precise identification does not matter for the \textit{ray grid} method. 
In practice the ray nearest to an edge of the active region is chosen. 
For quantities that span multiple layers, we mark them with superscript layer indices ($\in \{l,m,n\}$).

\begin{description}
  \item[{\(p^l\)}] the pitch vector for layer \(l\) gives the displacement perpendicular to the layer's rays and 
has magnitude that of the pitch separation of two neighboring rays.
  \item[{\(c^n\)}] the origin vector for layer \(n\) locates the center point of the specially identified ray $i=0$.
  \item[{\(r^{lm}_{ij}\)}] the crossing point of ray \(i\) from layer \(l\) and ray \(j\) from layer \(m\).
  \item[{\(r^{lm}_{00}\)}] the crossing point the specially identified zero rays of layers \(l\) and \(m\).
  \item[{\(w^{lm}\)}] a vector giving the displacement along the direction of a ray in layer \(l\) between the 
crossing points of that ray and two neighboring rays from layer \(m\).
\end{description}

The last two tensors, \(r^{lm}_{00}\) and \(w^{lm}\) can be calculated with simple vector arithmetic in a pair-wise manner among the \(N_l\) layers in the ray grid. 
The former is symmetric and both have undefined diagonals. 
These do require \(\mathcal{O}(N_l^2)\) operations where \(N_l\) is the number of layers and typically \(N_l = 5\). 
Given these two tensors, arbitrary crossing points of rays from two different layers can be written as,

\begin{equation}
r^{lm}_{ij} = r^{lm}_{00} + j w^{lm} + i w^{ml}.
\end{equation}

By exploiting the constant ray direction and pitch, this tensor replaces numerous calls to $sin()$, $cos()$ and $sqrt()$ functions with multiplication, addition and array indexing. 

The location of a crossing point $r^{lm}_{ij}$ along the pitch direction of a third layer $n$ may be written as,

\begin{equation}
p^{lmn}_{ij} = (r^{lm}_{ij} - c^n) \cdot \hat{p}^n.
\end{equation}
Expanding this in terms of \(r^{lm}_{ij}\) shows the number of unique vector operations needed to form this tensor is combinitoric in the number of layers (eg, $N=5$) and independent from the number of wires:

\begin{equation}
P^{lmn}_{ij} = r^{lm}_{00}\cdot \hat{p}^n + jw^{lm} \cdot \hat{p}^n + iw^{ml} \cdot \hat{p}^n - c^n \cdot \hat{p}^n.
\end{equation}

\noindent Finally, the typical use of this pitch location is to identify the nearest ray in plane $n$ by its index.  This index can be found simply as,

\begin{equation}
I^{lmn}_{ij} = floor(P^{lmn}_{ij}/p^n).
\end{equation}


\bibliographystyle{hunsrt}

\bibliography{dnn-roi}{}

\end{document}